# Paraxial design of four-component zoom lens system with fixed distance between focal points by matrix optics

ZICHAO FAN[1], ZHENGBO ZHU[2], SHILI WEI[2], YAN MO[2], AND DONGLIN MA[2, *]

[1]*MOE Key Laboratory of Fundamental Physical Quantities Measurement & Hubei Key Laboratory of Gravitation and Quantum Physics, PGMF and School of Physics, Huazhong University of Science and Technology, Wuhan 430074, China*
[2]*School of Optical and Electronic Information and Wuhan National Laboratory for Optoelectronics, Huazhong University of Science and Technology, Wuhan 430074, China*
*\*madonglin@hust.edu.cn*

**Abstract:** In this paper, we propose a systematic approach to design a four-component zoom system with fixed spacing between focal points based on matrix optics. Since the more complex model the higher degree of freedom it has, the task of determining the zoom trajectory is meaningful and challenging. The elements of the system matrix imply the working state of the optical system, and axial displacement equation for the desired zoom system are derived by restricting specific matrix elements. Properly selected trajectory of the particular component described by means of a parametric function can make the model become solvable explicitly. Then the paraxial design problem is transformed into the optimization of these parameters with regard to the merit functions encompassing the primary aberration terms, compactness, smoothness of the trajectories. We adopt Particle Swarm Optimization (PSO) algorithm to globally optimize the parameters to retrieve the optimum zoom trajectory in specific design criteria. The proposed method is demonstrated through two numerical examples under different configurations. The simulation results demonstrate that our proposed method can be a practical and powerful tool for paraxial design of complex multi-group zoom optical systems.

## 1. Introduction

Zoom optical systems are widely used in many areas as the variable focal length can meet the requirements in a variety of situations [1-3]. In some cases, we are much more concerned about keeping the two foci fixed instead of maintaining fixed conjugate distance. A zoom system with fixed focal planes is favorable in the field of optical information processing and machine vision [4-6]. Such zoom lenses are usually used in a 4-*f* system with variable magnification or as a part of a double-side telecentric lenses with variable magnification. It has been proved that the number of optical components for such a lens system always exceeds two [7]. In general, Gaussian brackets and purely algebraic methods [8] are the effective methods to perform a theoretical analysis of paraxial properties [9-11]. However, the mathematical derivation gets complicated for a multi-group zoom optical system, especially when the number of groups is greater than three. Miks presents a highly symmetric model to simplify the problem based on Gaussian brackets method, but the increase in the number of components doesn't significantly improve the zoom capability with this configuration [12]. Moreover, it is well known that the zoom system's performance is greatly determined by the paraxial structure, especially for the zoom capability. Therefore, a new analytical method deserves to be explored to achieve an optimized paraxial design of the described zoom system with a wide range of focal length variations.

Matrix optics method is a powerful tool in dealing with the initial configuration of optical systems. It is usually adopted for the task of determining the paraxial structure of complex optical systems. Kryszczynski has been trying to popularize this method in recent years and made significant contributions to the field [13-16]. He proposed a method based on the differentiation of system matrix to analyze the movements of the components in zoom systems. In matrix optics, a complex optical system can be described by means of a unitary quadratic matrix. Compared with other methods, the matrix description of optical systems is simpler and more general, as the elements of the system matrix are directly related to the first-order parameters of the optical system. Based on paraxial ray tracing, optical powers and spaces between components can be expressed by optical matrix and transfer matrix.

In this work, we focus on the problem of initial configuration construction for the four-component zoom optical system with fixed distance between focal points. By analyzing the paraxial properties of the zoom system, we obtain the characteristics of each element in the system matrix. With the help of system matrix, we have derived equations which enable us to calculate the required locations of individual components in the proposed system. The equation groups become solvable by presetting the trajectories of specific components to reduce the degrees of freedom. When the parameters of the trajectory equation are determined, the configuration of the zoom optical system is uniquely determined. In order to obtain reliable paraxial design, artificial intelligence algorithm is adopted to optimize its parameters. The optimization is performed in consideration of the total length, primary aberrations, and smoothness of zoom curve. The proposed method is demonstrated to be efficient and robust through the initial design of two typical examples. It is noted that our proposed method can also be a good guide for the initial design of any other complex optical system.

## 2. Theory of Matrix optics

### 2.1 Paraxial ray tracing and system matrix

First of all, we will consider tracing the marginal ray. We assume that the whole system is in the air ($n = n' = 1$). Figure 1 describes the paraxial ray tracing configuration based on the thin lens model of the whole optical system, where we apply the ray's paraxial angle ($u$) in each optical space as well as its height at each optical component ($y$) to represent the light path. According to the paraxial approximation and Gaussian imaging formula, we can obtain the following relations:

$$u = \frac{y}{-l}, \quad u' = -\frac{y}{l'}, \tag{1}$$

$$\frac{1}{l'} - \frac{1}{l} = \frac{1}{f}. \tag{2}$$

where $l$ is the object distance from lens to the object, $l'$ is the image distance from the lens to the image, and $f$ is the effective focal length of the lens.

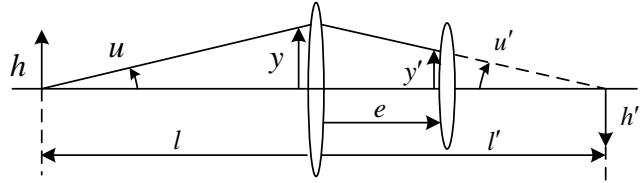

Fig. 1. Paraxial ray tracing model. $l$, $l'$: object distance and image distance respectively. $e$: distance from the current component to the next component. $u$, $u'$: paraxial angles consistent with the numeric apertures. $y$, $y'$: heights of the traced light ray at each optical component.

Then we can acquire the paraxial ray tracing formulas and rewrite the equations as a matrix form:

$$\begin{cases} u' = u - y\phi \\ y' = y + eu' \end{cases}, \tag{3}$$

$$\begin{cases} R = \begin{bmatrix} 1 & 0 \\ -\phi & 1 \end{bmatrix}, & \begin{bmatrix} y \\ u' \end{bmatrix} = R \cdot \begin{bmatrix} y \\ u \end{bmatrix} \\ T = \begin{bmatrix} 1 & e \\ 0 & 1 \end{bmatrix}, & \begin{bmatrix} y' \\ u' \end{bmatrix} = T \cdot \begin{bmatrix} y \\ u' \end{bmatrix}, \end{cases} \qquad (4)$$

$$\begin{bmatrix} y' \\ u' \end{bmatrix} = T \cdot R \cdot \begin{bmatrix} y \\ u \end{bmatrix}, \qquad (5)$$

where $\phi$ represents the optical power of the current optical element, $R$ is the optical power matrix and $T$ is the transfer matrix. The coordinates of characteristic paraxial rays at the boundaries can be described by means of these two quadratic matrices. Suppose that a thin lens system consists of $n$ components labelled as from 1 to n sequentially. The reference plane in the object space is labelled as 0 and the reference plane in the image space is labelled as $n+1$. Then we can get:

$$\begin{bmatrix} y_{n+1} \\ u_{n+1} \end{bmatrix} = T_n \cdot R_n \cdot T_{n-1} \cdot R_{n-1} \ldots T_1 \cdot R_1 \cdot T_0 \cdot \begin{bmatrix} y_0 \\ u_0 \end{bmatrix} = S \cdot \begin{bmatrix} y_0 \\ u_0 \end{bmatrix}, \qquad (6)$$

where $S$ is defined as the system matrix.

## 2.2 F-F′ system and system matrix

In this work, a zoom system with fixed distance between focal points is the primary object of study. The reference planes of the system are the front focal plane (FFP) and the back focal plane (BFP), which do not satisfy the conjugate relation. Such a system is called the F – F′ system. In case of the four-component F – F′ system shown in Fig. 2, two paraxial feature rays, the marginal ray and the chief ray, are traced separately. Based on the conjugate properties of optical systems, we can know that the angles $u_0 = \alpha_{n+1} = 0$ and the coordinates $H_0 = h_{n+1} = 0$. From the definition of the system matrix, we can obtain the following equation [16]:

$$\begin{bmatrix} 0 & H_{n+1} \\ -u_{n+1} & 0 \end{bmatrix} = \begin{bmatrix} S_1 & S_2 \\ S_3 & S_4 \end{bmatrix} \cdot \begin{bmatrix} h_0 & 0 \\ 0 & \alpha_0 \end{bmatrix}. \qquad (7)$$

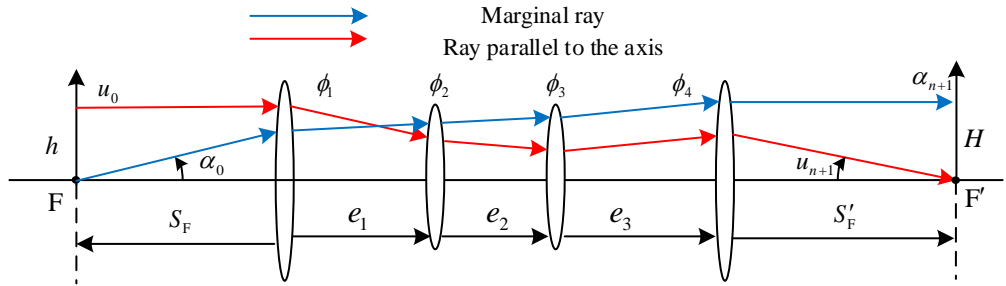

Fig. 2. Paraxial ray tracing model of a four-component F – F′ system. F, F′: front focal point and back focal point. $u_0$, $u_n$: ray slopes. $\alpha_0$, $\alpha_n$: paraxial angles. e: distance from the current component to the next component. $S_F$: distance from the first thin lens to FFP. $S'_F$: distance from the last thin lens to the BFP. $\phi$: optical power of current component. h, H: height of the traced ray at a specific surface.

According to the rules of matrix operation, we can acquire:

$$S_1 \cdot h_0 = 0 \; ; \; S_4 \cdot \alpha_0 = 0 \qquad (8)$$

$$S_2 \cdot \alpha_0 = H_{n+1} \; ; \; S_3 \cdot h_0 = -u_{n+1} \qquad (9)$$

From Eq. (8), it is clear that in case of an F – F′ optical system the selected matrix elements $S_1$ and $S_4$ should have the following values:

$$S_1=0 \quad ; \quad S_4=0 \tag{10}$$

From Eq. (9), we can derive that $S_2$ represents the effective focal length of the system, and $S_3$ is the opposite of the optical power of the entire system. Equation (10) ensures that the reference planes are always the focal planes of the optical system. In such a system, the system matrix $S$ should have the following form:

$$S=\begin{bmatrix} 0 & f \\ -\phi & 0 \end{bmatrix}. \tag{11}$$

### 2.3 Axial displacement equation

A zoom system with mechanical compensation should relocate its components to change the focal length while maintaining the location of both focal planes. Compared to traditional zoom system, we need another design degree of freedom to keep both focal planes stationary, because both focal planes are not conjugate with each other. Therefore, the simplest structure should consist of three independently moving elements. In our previous work, we derived the axial displacement equation for the three-component structure and presented an automated design method to achieve a paraxial design by means of Particle Swarm Optimization (PSO) algorithm [17-18]. However, the zoom capability of that system is limited to some extent, as increasing the zoom ratio leads to an unacceptable optical length. This problem might be solved by adding design freedoms, such as a four-component structure.

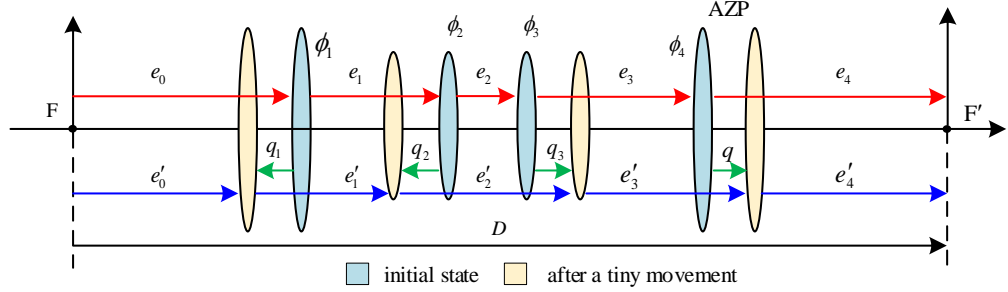

Fig. 4. The movement of the components

The zooming progress of the optical system is illustrated in Fig. 4, where the displacement $q$ is predetermined, and we label the corresponding component as the active zoom part (AZP). This creates two system matrices which can be derived from Eq. (6) before and after the movement separately. The system matrix $S_i$ of initial state is calculated as:

$$S_i = \begin{bmatrix} 1 & e_4 \\ 0 & 1 \end{bmatrix} \cdot \begin{bmatrix} 1 & 0 \\ -\phi_4 & 1 \end{bmatrix} \cdot \begin{bmatrix} 1 & e_3 \\ 0 & 1 \end{bmatrix} \cdot \begin{bmatrix} 1 & 0 \\ -\phi_3 & 1 \end{bmatrix} \cdot \begin{bmatrix} 1 & e_2 \\ 0 & 1 \end{bmatrix}$$
$$\cdot \begin{bmatrix} 1 & 0 \\ -\phi_2 & 1 \end{bmatrix} \cdot \begin{bmatrix} 1 & e_1 \\ 0 & 1 \end{bmatrix} \cdot \begin{bmatrix} 1 & 0 \\ -\phi_1 & 1 \end{bmatrix} \cdot \begin{bmatrix} 1 & e_0 \\ 0 & 1 \end{bmatrix}, \tag{12}$$

where $e_0$ is the distance from the front focal plane to the first surface, $e_4$ is the distance from the last surface to the back focal plane. Obviously, $e_0 = -S_F$ and $e_4 = S'_F$. Then a preset displacement $q$ is specified for the AZP, and other components need to compensate the shift of the focal planes caused by this displacement by varying the displacement of $q_1$, $q_2$ and $q_3$. To determine an increment of the matrix $\Delta S$ caused by movements of these components, the current transmission matrix can be obtained by Eq (4):

$$T_{14} = \begin{bmatrix} 1 & e_4 - q \\ 0 & 1 \end{bmatrix} \ ; \ T_{13} = \begin{bmatrix} 1 & e_3 - q_3 + q \\ 0 & 1 \end{bmatrix} ; \ T_{12} = \begin{bmatrix} 1 & e_2 - q_2 + q_3 \\ 0 & 1 \end{bmatrix} \ ;$$
$$T_{11} = \begin{bmatrix} 1 & e_1 - q_1 + q_2 \\ 0 & 1 \end{bmatrix} \ ; \ T_{10} = \begin{bmatrix} 1 & e_0 + q_1 \\ 0 & 1 \end{bmatrix}.$$
(13)

where $T_{in}$ is the *n*-th (*n*=0,1…4) transfer matrix after the *i*-th iteration. Substituting Eq. (13) into Eq. (6), the new system matrix $S_{i+1}$ can be obtained as:

$$S_{i+1} = \begin{bmatrix} 1 & e_4 - q \\ 0 & 1 \end{bmatrix} \cdot \begin{bmatrix} 1 & 0 \\ -\phi_4 & 1 \end{bmatrix} \cdot \begin{bmatrix} 1 & e_3 - q_3 + q \\ 0 & 1 \end{bmatrix} \cdot \begin{bmatrix} 1 & 0 \\ -\phi_3 & 1 \end{bmatrix} \cdot \begin{bmatrix} 1 & e_2 - q_2 + q_3 \\ 0 & 1 \end{bmatrix}$$
$$\cdot \begin{bmatrix} 1 & 0 \\ -\phi_2 & 1 \end{bmatrix} \cdot \begin{bmatrix} 1 & e_1 - q_1 + q_2 \\ 0 & 1 \end{bmatrix} \cdot \begin{bmatrix} 1 & 0 \\ -\phi_1 & 1 \end{bmatrix} \cdot \begin{bmatrix} 1 & e_0 + q_1 \\ 0 & 1 \end{bmatrix},$$
(14)

The stabilization conditions in the matrix form are implicit in Eq. (11), and the axial displacement equation can be written as:

$$\begin{cases} \Delta S = S_{i+1} - S_i \\ \Delta S[1,1] = 0 \ ; \ \Delta S[2,2] = 0 \ ; \ \Delta S[1,2] = \Delta f \end{cases}$$
(15)

The movements of the components $q_1$, $q_2$, and $q_3$ can be directly calculated by numerical solution of Eq. (15), and the effective focal length of the whole system will increase $\Delta f$ after an iteration due to the change of increment $\Delta q$.

So far, the optical information about the optical power $\phi_i$ of each component and optical space between two neighboring components are unknown. We can definitely configure the initial design referring to previous successful designs. However, an effective algorithm to search the optimal solution need to be developed, as the trajectory of AZP in the iterative process can be hardly determined via trial and error approach.

## 3. Optimization algorithm

### 3.1 Merit Function

The optimization algorithm for paraxial design of complex zoom systems highly depends on the choice of merit function reflecting the expected performance. Taking the preference of low aberration, smooth trajectory and compact structure into consideration, the final merit function is defined as:

$$T = w_A \cdot (\sqrt{S_{IS}^2 + S_{IIS}^2} + \sqrt{S_{IM}^2 + S_{IIM}^2} + \sqrt{S_{IL}^2 + S_{IIL}^2}) + w_L \cdot L + w_k \cdot \sum_{i=1}^{4} K_i,$$
(16)

where $w_A$, $w_L$, and $w_k$ are the weight factors corresponding to aberration, compactness, and smoothness separately, $S_I$ and $S_{II}$ are the Seidel aberration coefficients of different configurations at short focus, middle focus and long focus with subscripts of S, M, and L respectively. The basic equations used to calculate the primary Seidel aberrations are expressed by [19-20]

$$\begin{cases} S_I = -\sum A^2 \cdot y \cdot \Delta(\frac{u}{n}) \\ S_{II} = -\sum A\bar{A} \cdot y \cdot \Delta(\frac{u}{n}) \end{cases},$$
(17)

where *y* is the height of the marginal ray, *u* donates the slope of the marginal ray, *n* represents the refractive index, *A* is the refraction invariant of the marginal ray, and $\bar{A}$ is the refraction invariant of chief ray. They can be calculated as:

$$\begin{cases} A=n(yc+u) \\ \bar{A}=n(\bar{y}c+\bar{u}) \end{cases} \quad (18)$$

where $\bar{y}$ represents the height of the chief ray, and $\bar{u}$ is the slope of the chief ray. It allows easy imaging of characteristic rays in any optical surface using a matrix notation according to the Eq. (6). Heights and slopes of the incidence characteristic rays at different zoom position can be determined. $L$ defines the total optical length of the zoom system:

$$L = D - \min\{e_0\} - \min\{e_4\} \quad (19)$$

The smoothness of the cam curve is also evaluated, to confirm the kinematics of the moving components is reasonable. $K_i$ is defined as:

$$K_i = \sum K_{iN}$$
$$K_{iN} = \begin{cases} 0 & (e'_i(z_{N+1}) - e'_i(z_N) \le c) \\ |e'_i(z_{N+1}) - e'_i(z_N)| & (e'_i(z_{N+1}) - e'_i(z_N) > c) \end{cases} \quad (20)$$

where $c$ is the constant that represents an acceptable change in zooming speed. Since we solve for multiple discrete configurations of the zoom system, the first order derivative of the distance $e_{iN}$ is in respect to $q$ can be replaced with:

$$e'_i(z_N) = \frac{e_{iN}(q_N) - e_{iN-1}(q_{N-1})}{q_N - q_{N-1}}. \quad (21)$$

### 3.2 Particle swarm optimization algorithm

Based on the analysed results, the paraxial performance of zoom system is determined by the parameters of AZP trajectory function. The paraxial design of the zoom system is transformed into the multipeak optimization problem with several variables. With the advantages of high efficiency, rapid speed of convergence and strong capability of global search, the PSO algorithm is a simple and effective way to implement the optical system design [21-26]. Among the applications recorded in these literatures, PSO algorithm can not only be used for aberration optimization, but also for initial structure retrieval. As a result, PSO algorithm is adopted as a tool to deal with optimization problem.

Necessary information about PSO algorithm is provided to illustrate our design approach. More details on the algorithm can be found in refs. [17-18]. In the PSO, a group of N particles are randomly assigned in the search space $R^S$, and the position of each particle is denoted by $p_i$ (i=1, 2, …, N). The whole set of particles called swarm are assumed as the P = {$p_1$, $p_2$, …, $p_k$}, where k represents the current iteration of the group. At each step, $p_i$ has a position vector $\mathbf{x}_i = (x_{i1}, x_{i2}, …, x_{iS})$ and a velocity vector $\mathbf{v}_i = (\upsilon_{i1}, \upsilon_{i1}, …\upsilon_{iS})$ to describe the attributes of the particle. $\mathbf{p}_{ibest}$ stores the best position that the $i$-th particle has achieved so far, and $\mathbf{g}_{best}$ is the best position of the swarm in its history. Then the particle updates its status through the rules of combining aspects of $\mathbf{p}_{ibest}$, $\mathbf{g}_{best}$, and some random perturbations:

$$\mathbf{v}_i^k = w \cdot \mathbf{v}_i^{k-1} + w_1 \cdot ran \cdot (\mathbf{p}_{ibest}^k - \mathbf{x}_i^k) + w_2 \cdot ran \cdot (\mathbf{g}_{best} - \mathbf{x}_i^k)$$
$$\mathbf{x}_i^{k+1} = \mathbf{x}_i^k + \mathbf{v}_i^k \quad (22)$$

where $w$ is the inertia coefficient, $w_1$ and $w_2$ are acceleration coefficients, and ran generates uniform random factors in the [0, 1] interval. Adopting a linear decreasing model of the inertia weight makes the algorithm possess capability of exploring the entire search space in the early iterations and the capability of convergence to optima in the later stages. The inertia weight is usually determined in accordance to the formulae:

$$w = w_{max} - \frac{w_{max} - w_{min}}{iter} \cdot k \quad (23)$$

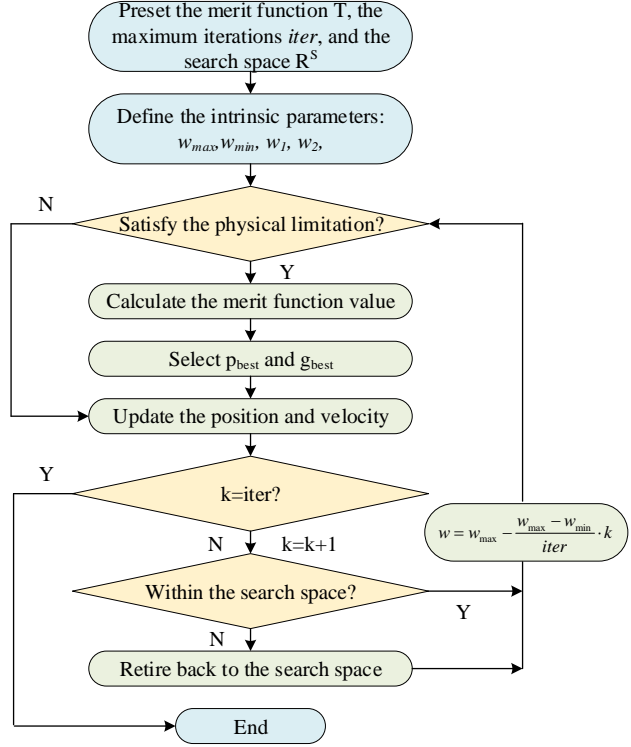

Fig. 5. Flow chart of design process

With the merit function defined in Eq. (16), we can apply the PSO algorithm to find the optimal initial paraxial design configuration. We set the trajectory function of AZP as $q(x_0, x_1, x_2,..x_S)= x_0 i^0+x_1 i^{1/2}+x_2 i+ x_3 i^{3/2}+…$, where $i$ is the cycle parameter. The variable parameters are represented by particles with the position vector $\mathbf{x}_i= [x_0, x_1, x_2…, x_S]$. The general process of our modified PSO algorithm is depicted in Fig. 5. The swarm is initialized under the rules mentioned above and the intrinsic parameters are set. Afterward, according to Eqs. (17)-(21) and some basic knowledge of geometrical optical, calculate the evaluation function value of each particle, so as to determine $\mathbf{p}_{ibest}$ and $\mathbf{g}_{best}$ at this time. Then, the particles are updated according to the rules described in Eq. (22). If the particle escapes the search space, define a compensation mechanism that allows a small exploration around the original location while pulling the particle back into the search space.

$$\mathbf{x}_i^{k+1} = \mathbf{x}_i^k - r_b \cdot \mathbf{v}_i^k, \tag{24}$$

where $r_b$ defines random numbers in the range [−1.1, 1.1]. When the pre-determined the maximum number of iterations is achieved, the optimization is terminated. If the obtained paraxial design cannot meet the requirements of design, the dimension of search space is added to increase the degree of freedom of design.

## 4. Design results and discussion.

### 4.1 Example of an 8X PNNP zoom system

According to the mathematical models established above, we implement the method in Matlab to design an 8X zoom system with four-components. When calculating the merit function value of the position of each particle, it is necessary to solve the zoom trajectory composed of multiple discrete positions corresponding to it. Therefore, considering the running speed of the program, the selection of these positions should not be too intensive. Here we set 120 sampling positions to discretize the whole zoom trajectory.

Considering the optical power contribution of each component to the optical system and symmetry of the system, the power combinations such as P-N-N-P and P-N-P-P are usually preferred choices for a four-component zoom system, where P represents positive optical power of the component, and N represents negative optical power of the component. These configurations have been tried in early testing, and the results show that we can obtain reasonable solutions for subsequent optimization in both cases. First, we design an optical system with N-P-P-N structure. The distribution of optical power of each component and the space distance between each component are located in Table 1. The wide-angle configuration is chosen as the starting point. The fourth component is selected as the AZP, and the other components serve as compensators to satisfy the zoom conditions.

Table 1. The design data of starting point

|  | $f_0$=26mm |  | D=70.01mm |  | AZP: the 4$^{th}$ component |
|---|---|---|---|---|---|
| Distance (mm) | $e_0$ | $e_1$ | $e_2$ | $e_3$ | $e_4$ |
|  | -4.54 | 1.78 | 7.00 | 18.97 | 50.81 |
| Optical power (mm$^{-1}$) | $\phi_1$ |  | $\phi_2$ | $\phi_3$ | $\phi_4$ |
|  | 0.0133 |  | -0.0250 | -0.0346 | 0.0440 |

The detailed parameters for the PSO algorithm are also listed in Table 2. The particle swarm size is set as $N$=100 to ensure that there are enough particles that meeting the physical restriction at the initial stage of optimization. The value of parameters $w_A$, $w_L$, $w_k$ are set as 25, 1 and 3 respectively. In the paraxial ray tracing, the aperture of the optical system is 10 mm and the FOV is 1mm. The kinetics of the four moving components is obtained after optimization, which are shown in Table 3. The loci of all the four components for the system are as shown in Fig. 6.

Table 2. Parameters of the PSO algorithm

| R$^S$(mm$^{-1}$) |  | $x_0 \in$ [-1, -0.1] |  | $x_1 \in$ [0, 0.2] | $x_2 \in$ [0.001, 0.1] |  | $x_3 \in$ [-0.01, 0] |  |
|---|---|---|---|---|---|---|---|---|
|  | $w$ |  | $w_1$ | $w_2$ | Merit function |  | N | c | $iter_{max}$ |
| PSO | $w_{max}$ | $w_{min}$ | 0.6 | 0.8 | $w_A$ | $w_L$ | $w_k$ | 100 | 2 | 100 |
|  | 1.2 | 0.2 |  |  | 25 | 1 | 3 |  |  |  |

Table 3. The design data of 8X zoom lens (Unit: mm)

|  | $x_0$=-0.47382 | $x_1$=0.05205 | $x_2$=0.009055 | $x_3$=-0.00019 |  |
|---|---|---|---|---|---|
| $f_0$ | $e_1$ | $e_2$ | $e_3$ | $e_0$ | $e_4$ |
| 26 | 1.7800 | 7.0000 | 18.9654 | -4.5436 | 50.8129 |
| 72 | 32.8260 | 14.9365 | 10.6526 | -39.8414 | 55.4409 |
| 118 | 40.8260 | 33.7281 | 5.9367 | -59.6891 | 53.2130 |
| 164 | 43.6688 | 54.7818 | 3.1721 | -72.7646 | 45.1565 |
| 208 | 44.7975 | 75.7888 | 1.4374 | -80.5983 | 32.5892 |

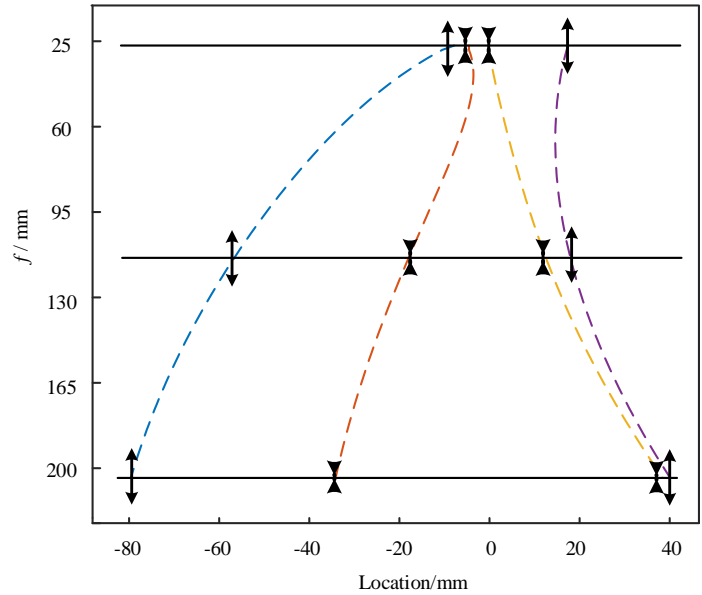

Fig. 6 Loci of the four components of the 8X P-N-N-P zoom system

In order to verify that the first-order design results obtained by this method can be effectively implemented in the actual lens system, the paraxial design is treated as the starting point for subsequent optimization in optical software Zemax. Necessary lens splitting is carried out for each component of the system to maintain the original optical power and increase the degrees of freedom that can be optimized. While the field of view is gradually expanding, the curvature radius of each surface is sensibly adjusted to make the beam constrained properly by the optical system. In the process of optimization, the coordinates of characteristic rays on the front focal plane are constrained under multiple zoom structures to ensure that the focal planes remain stationary.

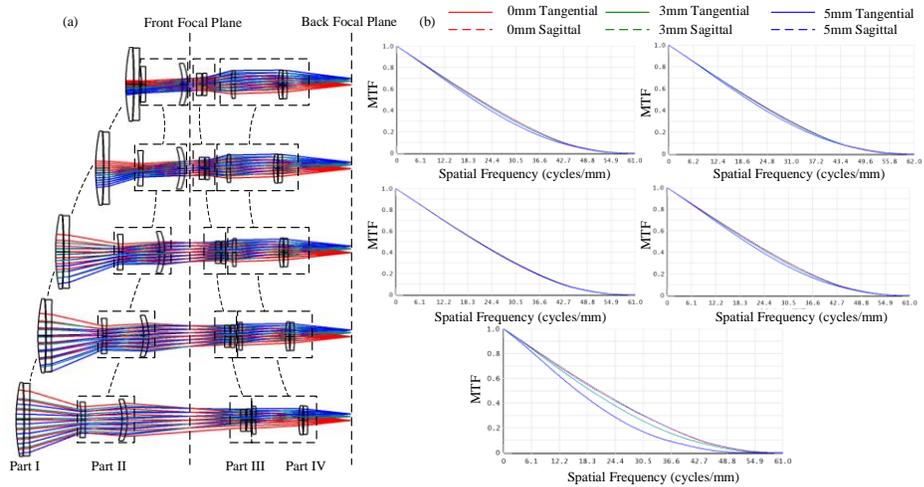

Fig. 7. Optimized result with actual lens: (a) layout of designed zoom system (b) MTF plots for the five zoom configurations

By means of optimizing the optical system to reduce the aberrations introduced by actual lenses and compensating the slight deviation of paraxial zoom trajectory, a four-component zoom system with two fixed foci has been successfully gained. The working wavelength of the optical system ranges from 3.7μm to 4.8μm, and the aperture of the system is F/# =4.5. The effective focal length of the system ranges from 26mm to 208mm, with a maximum optical length of 259.22mm for the configuration with long focal length. Figure 7(a) presents the layout of the designed zoom system at five different zoom ratios. In Fig. 7(b), the corresponding modulation transfer function (MTF) curve is given as the evaluation standard of imaging quality. Since the merit function well reflects the characteristics of the zoom system, the proposed paraxial design method plays a key role in the subsequent optimization.

*4.2 Example of an 8X PNPP zoom system*

Compared to previous configuration, P-N-P-P is more competitive in optical length. In addition, the rays in the aperture are more confined to the paraxial region, so the lens size is a bit smaller. The disadvantage is that more investment is needed in aberration control due to the slightly poor symmetry of the system. Therefore, in the paraxial design stage, we increase the weight of coefficient $w_A$ to control aberration term, and the subsequent optimization method in optical design software is to introduce aspheric surface to balance the aberration.

**Table 4. The design data of starting point**

|  | $f_0$=24mm | D=69.48mm | AZP: the 4$^{th}$ component | | |
|---|---|---|---|---|---|
| Distance (mm) | $e_0$ | $e_1$ | $e_2$ | $e_3$ | $e_4$ |
|  | -6.32 | 1.87 | 9.77 | 21.67 | 47.81 |
| Optical power (mm$^{-1}$) | $\phi_1$ | $\phi_2$ | $\phi_3$ | $\phi_4$ | |
|  | 0.0086 | - 0.0820 | 0.0177 | 0.0422 | |

The data of the starting point is shown in Table 4, including the distance between the neighbouring components, the optical power of each component, and the choice of AZP element. In order to make the particles converge to the optimal position more quickly, the search space can be adjusted according to the solution retrieval experience. This ensures that there are enough available particles wandering in the search space to provide information for the swarm to make a global search. After performing plenty of numerical tests, we finally obtain the parameters used in the PSO algorithm for the first-order design of the special zoom system as listed in Table 5. Since the solvable region is narrower than the configuration in the previous chapter at the same zoom ratio, more particles are required to be put into the search space. Moreover, the coefficient $w_2$ is raised to guide other particles to move to the solvable region more quickly. Other parameters are basically the same as those in Section 4.1.

**Table 5. Parameters of the PSO algorithm**

| $R^S$(mm$^{-1}$) | | $x_0 \in$[-1, -0.2] | | $x_1 \in$[-0.2, 0] | $x_2 \in$[0, 0.1] | | $x_3 \in$[-0.05, 0.05] | | |
|---|---|---|---|---|---|---|---|---|---|
| PSO | | w | | $w_1$ | $w_2$ | Merit function | | N | c | $iter_{max}$ |
|  | $w_{max}$ | $W_{min}$ | 0.6 | 1 | $w_A$ | $w_L$ | $w_k$ | 150 | 2 | 100 |
|  | 1.2 | 0.2 | | | 20 | 0.5 | 2 | | | |

After 100 iterations of optimization by the modified PSO algorithm, a globally optimum initial design for the special 8X P-N-P-P zoom system is obtained, whose focal length ranges from 24mm to 196mm. The final detailed first-order design result is shown in Table. 6, and the loci of the four components of zoom system are illustrated in Fig. 8. Aspheric surfaces are used in aberration correction as paraxial design components are gradually replaced by actual lenses. These aspheric surfaces are placed on the surface near the aperture and the image plane, which are mainly used to correct spherical aberdration and field aberrations such as distortion and coma respectively. After repeated optimization, the layout of the final design and MTF at five different focal lengths are provided in Fig. 9.

Table 6. The design data of 8X zoom lens (Unit: mm)

|  | $x_0$=-0.31 | $x_1$=-0.14 | $x_2$=0.042 | $x_3$=-0.00027 |  |
|---|---|---|---|---|---|
| $f_0$ | $e_1$ | $e_2$ | $e_3$ | $e_0$ | $e_4$ |
| 24 | 8.2000 | 7.9000 | 11.9000 | -6.3299 | 47.8139 |
| 64 | 54.8566 | 8.3988 | 1.4040 | -49.2080 | 54.0325 |
| 106 | 77.3102 | 5.7258 | 3.7994 | -67.8341 | 50.4826 |
| 148 | 93.5211 | 3.4517 | 6.3707 | -68.2895 | 34.4300 |
| 192 | 107.7810 | 0.2354 | 10.5828 | -52.5856 | 3.4703 |

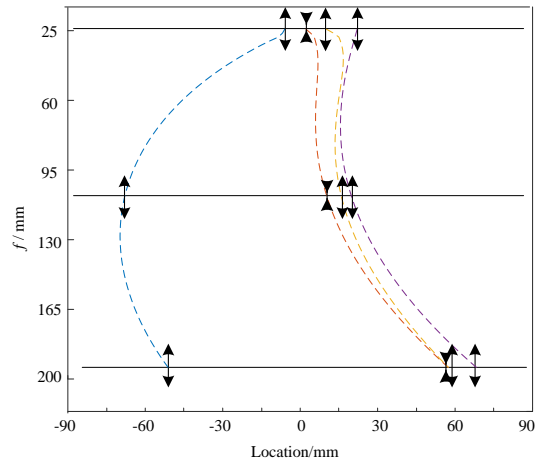

Fig. 8 Loci of the four components of the 8X P-N-P-P zoom system

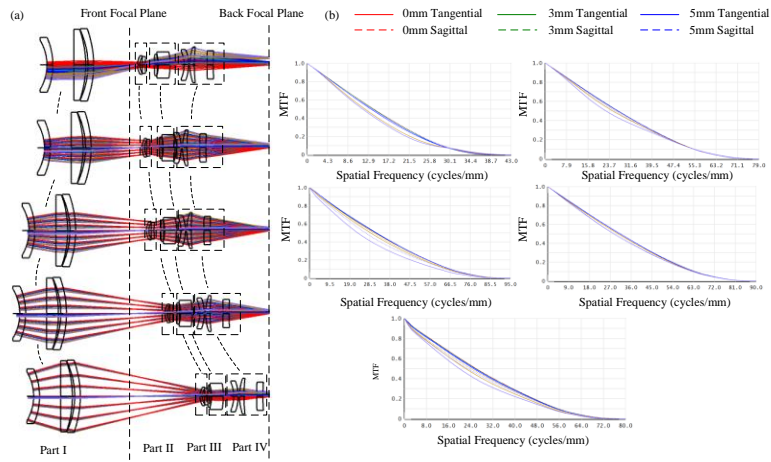

Fig. 9. Optimized result with actual lens: (a) layout of designed zoom system (b) MTF plots for the five zoom configurations

From the simulation results, the paraxial and actual zoom trajectories of the system are smooth and the aberrations are well corrected. Compared to our previous design [18], increasing the number of components inside the optical system can help achieve a relatively higher zoom ratio. A more accurate merit function reflecting the design requirements makes our paraxial design plays an important guiding role in the optimization of the actual system. The design examples illustrate that the initial design using our method can be used as a reliable starting point for subsequent optimization.

## 5. Conclusion

A systematic approach using matrix optics for the paraxial design of an F- F' zoom lens system is presented in this paper. Compared with the traditional analysis method, the matrix optics method is more concise to deal with the problem of multi-component zoom system, because the entire optical system can be regarded as a black box when the axial displacement equation is constructed. Then the kinematics inside the black box is analyzed by adjusting one of the components named AZP. Presetting the trajectory of AZP during zooming makes the axial displacement equation can be solved directly. Under this model, a global optimization based on the PSO algorithm has been created, allowing optimal paraxial design to be automatically retrieved for the specific zoom system. The experimental results show that the obtained the paraxial design can be successfully translated into practical zoom optical lenses by means of optimization skills that ordinary optical designers can handle.

As expected, the design degrees of freedom provided by the four-component system were successfully translated into the zoom capability of the practical system. If the design allows a larger optical length and more aspheric or freeform surfaces, the method can achieve a larger zoom ratio design. Moreover, compared with the traditional trial and error, this method significantly improves the efficiency and does not rely too much on the experience of the designer. This automated design method obviously provides designers with a simpler and more efficient tool for designing complex zoom systems. It not only ensures that the axial displacement equation can be solved accurately, but also retains the optimization potential brought by the extra degree of freedom as the number of components increase. This highly formalized approach provides clarity of thought for designers. In addition, it can also apply to ordinary zoom lens systems or other types of special zoom systems.

## Funding

National Natural Science Foundation of China (61805088); Science, Technology, and Innovation Commission of Shenzhen Municipality (JCYJ20190809100811375); Key Research and Development Program of Hubei Province (2020BAB121); Fundamental Research Funds for the Central Universities (2019kfyXKJC040); Innovation Fund of WNLO.

## Disclosures

The authors declare no conflicts of interest.